\documentclass{article}

\usepackage[preprint]{neurips_2026}

\usepackage[utf8]{inputenc}
\usepackage[T1]{fontenc}
\usepackage{hyperref}
\usepackage{url}
\usepackage{booktabs}
\usepackage{array}
\newcolumntype{L}[1]{>{\raggedright\arraybackslash}p{#1}}
\usepackage{amsfonts}
\usepackage{amsmath}
\usepackage{nicefrac}
\usepackage{microtype}
\usepackage{xcolor}
\usepackage{graphicx}
\graphicspath{{figures/}}

\title{TNFlow: Amortized Posterior Inference for Trans-Neptunian Object Surface
       Composition}

\author{%
  Agastya Gaur \\
  Department of Astronomy, University of Illinois Urbana-Champaign, Urbana, IL 61801 \\
  SETI Institute, 339 Bernardo Avenue, Suite 200, Mountain View, CA 94043 \\
  \texttt{gaur4@illinois.edu} \\
  \AND
  Cristina Dalle Ore \\
  Carl Sagan Center, SETI Institute, 339 Bernardo Avenue, Suite 200, Mountain View, CA 94043 \\
  \texttt{cdalleore@seti.org} \\
  \AND
  Alessandra Ricca \\
  NASA Ames Research Center, MS 245-6, Moffett Field, CA 94035-1000 \\
  Carl Sagan Center, SETI Institute, 339 Bernardo Avenue, Suite 200, Mountain View, CA 94043 \\
  \texttt{alessandra.ricca-1@nasa.gov}
}

\begin{document}
\maketitle

\begin{abstract}
We present TNFlow, a transformer and normalizing flow architecture for inferring the surface composition of Trans-Neptunian Objects (TNOs) from their reflectance spectra. TNFlow is trained on synthetic spectra generated by the Shkuratov radiative transfer model to act as its inverse. TNFlow takes ${\sim}$0.7\,s to invert one spectrum on a single CPU core, returning a multimodal posterior over simplex-valid compositions and grain sizes. On synthetic spectra, the highest-weight mode achieves a mean total-variation distance of 0.149 from ground truth on the test split, and the model generalizes well to unseen combinations of known components. Qualitative tests on real JWST spectra show blindness or bias towards some materials. We suggest this could be attributed to either simulator fidelity or the training set.
\end{abstract}

\section{Introduction}

Trans-Neptunian objects (TNOs) are among the most primitive surfaces accessible to observation in the solar system. Frozen since their formation, they retain volatile ices whose abundances encode the history of the protoplanetary disk. Recovering the composition of these ices, what materials are present in what proportion and grain size, is a direct link to the formation conditions of the early solar system.

Surfaces are measured spectroscopically, producing a reflectance spectrum: how much light the surface reflects as a function of wavelength. Each ice absorbs at characteristic wavelengths, producing dips whose depths and shapes can, in principle, be used to infer composition.

Inference is difficult, however, because the composition-to-spectrum mapping is many-to-one. Spectra can be generated from a composition using a radiative transfer model (RTM) \citep{hapkeBidirectionalReflectanceSpectroscopy1981,hapkeBidirectionalReflectanceSpectroscopy1984,hapkeBidirectionalReflectanceSpectroscopy1986,shkuratovModelSpectralAlbedo1999}, but a standard method for RTM inversion remains an open question.

Fast, degeneracy-aware RTM inversion is the first step towards a large-scale model to infer compositions from real-life spectra. This work introduces TNFlow: a transformer and normalizing flow based architecture trained for TNO surfaces using the Shkuratov RTM \citep{shkuratovModelSpectralAlbedo1999}. Once trained, TNFlow yields a full posterior density of possible compositions in under a second, making survey-scale analysis practical. Since TNFlow is trained entirely on simulator output, we also test the sim-to-real gap qualitatively using JWST measurements (\autoref{app:jwst}). Code and data are available at \url{https://github.com/agastiyo/TNFlow-companion}.

\section{Related Work}
\label{sec:related}

\textbf{Other solutions.}
Traditionally, spectra have been analyzed manually. An analyst assembles candidate mixtures based on features visible in the spectrum, runs the forward model, and selects solutions by eye \citep{cruikshankPlutoEvidenceMethane1976,cruikshankIcesSurfaceTriton1993,merlinSurfaceCompositionPhysical2010,barucci50000QuaoarSurface2015}. This is physically grounded and needs no training corpus, but is a subjective process, costs hours to days per object, and doesn't formally handle uncertainty.

MCMC inversion of an RTM \citep{schmidtRealisticUncertaintiesHapke2015} is automated and returns a multimodal posterior, and it documents degeneracy in this class of inversion directly. However, the forward model is evaluated inside the sampling loop, which makes the cost per object large. Linear spectral unmixing \citep{liuEndmemberIdentificationSpectral2016} runs in seconds, but its output is a single point estimate with a local Gaussian error bar.

Finally, learned inverse models similar to TNFlow only exist in adjacent domains \citep{marquez-neilaSupervisedMachineLearning2018,vasistNeuralPosteriorEstimation2023,cranmerFrontierSimulationbasedInference2020,papamakariosNormalizingFlowsProbabilistic2021}. Population-scale work on TNOs stops at qualitative taxonomy rather than abundances \citep{barucciTaxonomyCentaursTransNeptunian2005,hainautColoursMinorBodies2012,pinilla-alonsoJWSTDiSCoTNOsPortrait2025}.

No existing method for RTM inversion is directly comparable on accuracy to TNFlow except for MCMC, which is out of the scope of this paper (\autoref{app:comparison}). Therefore, we claim no superiority in accuracy over any other method. Running time is compared against MCMC in \autoref{sec:results}.

In summary, the key gap is a method that is fast enough for survey-scale use, returns a full posterior over compositions, and can be evaluated entirely offline once trained. The remainder of this paper describes TNFlow's approach to filling that gap.

\section{Method}

TNFlow is a normalizing flow model that learns to invert an RTM for TNO surface composition. Specifically, TNFlow uses a conditional neural spline flow (CNSF) \citep{durkanNeuralSplineFlows2019a}, an extension of a normalizing flow that conditions the mapping on an input, so that each input yields its own posterior distribution.

TNFlow has four components: a tokenizer, a Transformer \citep{vaswaniAttentionAllYou2017} context encoder, and two CNSFs, one for composition and one for grain size. The model has 1,463,650 trainable parameters and outputs simplex-valid compositions and grain sizes over $K$ materials.

\textbf{Tokenizer.} Raw reflectance spectra are fed into the tokenizer as $N$ channels of reflectance-wavelength pairs. Reflectance values are normalized to the mean reflectance in a window around $0.9 \mu$m, and the wavelength is positionally encoded with 6 Fourier pairs. The normalized reflectance and positionally encoded wavelength are concatenated to form $N$ $13$-dimensional tokens.

\textbf{Context Encoder.} The $N$ tokens are passed through a Transformer encoder with 4 multi-head attention layers, each with 8 heads and operating on a 128-dimensional embedding. The output of the multi-head attention layers is attention-pooled with learned weights to produce a single context vector of dimension 128.

\textbf{Composition CNSF.} The context vector is passed to a $(K-1)$-dimensional CNSF with 4 transforms, each with 16 bins and two hidden dimensions of size 128. The CNSF outputs a posterior distribution over the proportions of all candidate components. Raw draws are then prescaled by a factor of 2.5\footnote{Required so that inverse stick-breaking outputs during training remain within the CNSF's learnable range.} and passed through a stick-breaking transformation, to ensure every sample is a valid composition by construction.

\textbf{Grain CNSF.} This is run once per component holding more than $0.01$ proportion in a given flow draw\footnote{Components at or below $0.01$ proportion are treated as absent.}. The context vector, the full composition, and a 16-dimensional learned embedding of the current component condition a 1-dimensional CNSF with 2 transforms, each with 8 bins and two hidden dimensions of size 64. This flow yields a $K$-dimensional grain-size posterior without a $K$-dimensional flow.

\textbf{Training.} Each spectrum in the training set has a target composition and a target grain size per present component. The model is trained end-to-end on the sum of two negative log-likelihoods, weighted equally: the composition CNSF scores the target composition, and the grain CNSF scores the target grain sizes, standardized as $\log_{10}$. The grain CNSF is teacher-forced, conditioned on the ground-truth target composition rather than a draw from the composition CNSF\footnote{This is also why splits are keyed by composition group (\autoref{sec:data})}. The grain CNSF is masked to predict grain sizes only for components that are present in the target. Both terms backpropagate through the shared encoder. To narrow the gap between simulator output and real observations, synthetic spectra are augmented each epoch with multiplicative and additive Gaussian noise (\autoref{app:noise}). We optimize with Adam at a learning rate of $10^{-3}$ with plateau decay on a single GPU for approximately 8 hours (\autoref{app:compute}). Our reported checkpoint and plateau schedule is selected using the held-out test split that \autoref{sec:results} reports on. However, selection is over epochs only and uses a criterion distinct from the reported metrics. During evaluation, these spectra also had a different noise realization than during training.

\textbf{Inference.} No extra noise is added during inference. A spectrum is tokenized and encoded once, yielding the context vector that conditions both flows. Drawing further samples requires no re-encoding. For each draw, a composition is sampled from the composition CNSF and also conditions the grain CNSF for its own grain sizes. One draw yields a $K$-dimensional composition and a $K$-dimensional grain vector. $n$ draws give two $n \times K$ matrices. Marginals and credible intervals follow directly, and distinct solutions are recovered by fitting a BIC-GMM to the composition draws.

\section{Data}
\label{sec:data}

TNFlow is trained entirely on synthetic data: 463,275 reflectance spectra (0.35--5\,\textmu m) generated with the Shkuratov RTM \citep{shkuratovModelSpectralAlbedo1999}. Each spectrum is an intimate mixture of up to seven materials drawn from a library of $K{=}20$ candidate components covering the main ice and refractory families expected on TNO surfaces. The corpus is assembled from two sources described in \autoref{app:components}: a pre-existing collection of Shkuratov bestfit spectra (163,285 spectra, 1--4 components) and a purpose-built extension (299,990 spectra, 2--7 components) generated with CANA \citep{cana2018} and SDOC\footnote{\url{https://github.com/cana-asteroids/sdoc}}.

\textbf{Splits.}
14.7\% of spectra were byte-identical to another spectrum: same components and proportions, differing only in grain size. Naively splitting the data into training and testing sets would risk leakage, so we split by composition group: every spectrum is keyed by (component set, proportions) and all rows sharing a key are assigned to one split. Eight component sets are held out entirely as out of distribution (OOD), which refers to unseen combinations of the $K=20$ training components, not novel materials. The remainder is partitioned 80/20, giving 370,210 training, 92,691 test, and 374 OOD spectra.

\textbf{Real Spectra.}
Informal analysis on real spectra is given in \autoref{app:jwst} and uses data from the JWST DiSCo-TNOs survey \citep{pinilla-alonsoJWSTDiSCoTNOsPortrait2025}.

\section{Results}
\label{sec:results}

We evaluate TNFlow on all 374 OOD spectra and on 5,000 randomly sampled test spectra. All spectra are noised in the same way as the training spectra. For each spectrum, we draw 500 posterior samples and fit a BIC-GMM to the draws to identify distinct modes. The BIC-GMM returns all found modes, their weights, and the uncertainty of each mode. The top mode is the one with the highest weight, and the best mode is the one closest to the target composition in total-variation distance.

\begin{figure}[t]
  \centering
  \includegraphics[width=\linewidth]{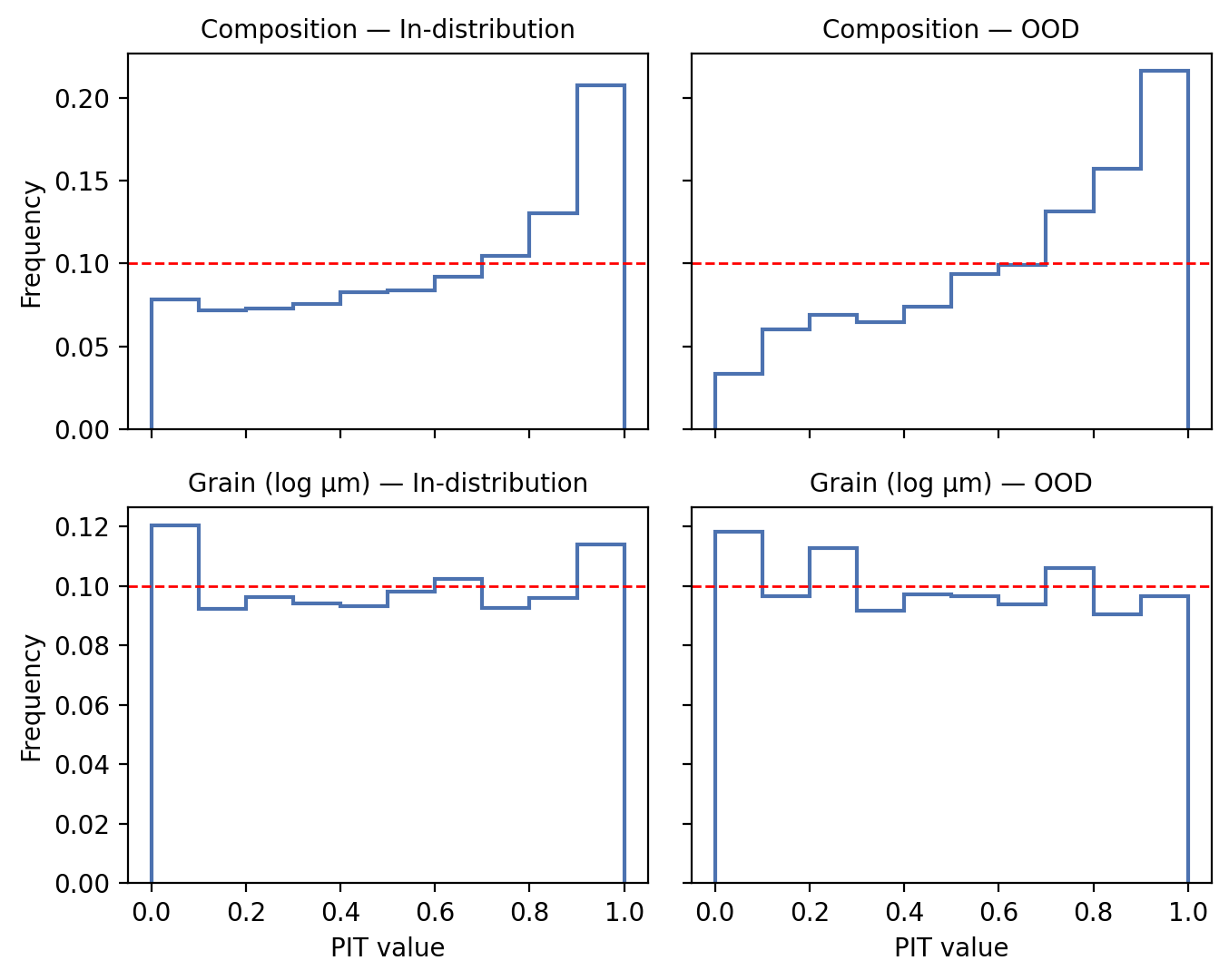}
  \caption{Probability integral transform (PIT) histograms for composition (top) and
           grain size (bottom), on the test (left) and OOD (right) splits.
           A uniform histogram (dashed) indicates calibration. Composition intervals are
           biased in both splits, with a right-skewed ramp indicating systematic underprediction that worsens OOD; grain
           intervals are well calibrated throughout.}
  \label{fig:pit}
\end{figure}

We measure composition accuracy in total-variation distance (TV). Grain accuracy is measured in dex ($\log_{10}$ error). We also report the fraction of spectra for which the top mode is also the best mode, which measures how often the highest-weight mode is also the most accurate. Accuracy results are summarized in \autoref{tab:results}. Calibration of the posterior is measured using probability integral transform (PIT) histograms on true components. A PIT histogram maps each true value to its percentile within the predicted distribution. If the posterior is well-calibrated, the PIT histogram is uniform. We report PIT histograms for both composition and grain size in \autoref{fig:pit}. We also report posterior sharpness: the mean pairwise Euclidean distance between draws.

\textbf{Cost.} TNFlow takes ${\sim}0.7$\,s per object on a single CPU core (\autoref{app:compute}). Training and corpus generation are one-time offline costs amortized over every subsequent object. For scale, \citet{schmidtRealisticUncertaintiesHapke2015} report Hapke MCMC retrievals of order $10^6$ forward-model evaluations taking 1\,min to 1\,h per object on a comparable CPU, with an accelerated variant giving a further ${\sim}100{\times}$ speedup.

\begin{table}[t]
  \caption{Per-solution accuracy on the test and group-holdout OOD splits
           (500 posterior draws per query, joint composition + grain BIC-GMM mode search).
           Top mode is the highest-weight mode; best mode is the mode closest to the
           target composition in total-variation distance. All $\pm$ values are one standard deviation.}
  \label{tab:results}
  \centering
  \begin{tabular}{lcc}
    \toprule
    Metric & Test ($n{=}5{,}000$) & OOD ($n{=}374$) \\
    \midrule
    Top-mode comp.\ TV      & $0.149 \pm 0.125$ & $0.156 \pm 0.091$ \\
    Best-mode comp.\ TV     & $0.121 \pm 0.098$ & $0.132 \pm 0.080$ \\
    Top-mode grain (dex)    & $0.242 \pm 0.212$ & $0.279 \pm 0.162$ \\
    Best-mode grain (dex)   & $0.253 \pm 0.201$ & $0.283 \pm 0.161$ \\
    Top mode = best (\%)    & 67.7  & 60.4  \\
    Mean mode count          & 2.50  & 2.64  \\
    Multimodal (\%)          & 88.9  & 91.7  \\
    Posterior sharpness      & $0.203 \pm 0.106$ & $0.228 \pm 0.066$ \\
    \bottomrule
  \end{tabular}
\end{table}

\section{Discussion}
\label{sec:discussion}

For test spectra, TNFlow recovers the top mode with a relatively accurate mean composition TV ($0.149 \pm 0.125$). Grain error remains around a factor of two of truth ($0.242 \pm 0.212$ dex). The best mode is slightly better in composition, but its grain error is slightly worse. This is most likely due to the best mode being chosen on composition distance. Performance on the OOD split degrades slightly, but remains close to the test results. This indicates that TNFlow retains similar performance on unseen combinations of known components.

The model's outputs are also multimodal. In 88.9\% of test and 91.7\% of OOD spectra, the model finds more than one distinct solution. The mean number of modes is 2.50 for test and 2.64 for OOD spectra. In 67.7\% of test spectra and 60.4\% of OOD spectra, the top mode is also the best mode. In the remainder, the model did not weight the most accurate solution highest. The relatively accurate top-mode TV distance indicates these may also be close solutions, but establishing that all found modes are genuine degenerate solutions requires resynthesizing each mode through the forward model, which we leave to future work.

The slight degradation in accuracy and increased multimodality on the OOD split points to a healthy increase in uncertainty as the model is asked to extrapolate beyond its training distribution. Posterior sharpness moves in the same direction: draws are more widely spread on the OOD split ($0.228$) than on the test split ($0.203$).

The grain PIT histograms are well-calibrated. The composition PIT shows a right-skewed ramp, indicating systematic underprediction that worsens OOD. This is consistent with the model occasionally placing mass on extra components: because compositions are simplex-valid by construction, any mass assigned to a spurious component necessarily reduces the proportions of the true ones. The best mode's lower TV ($0.121$ vs.\ $0.149$ on test) confirms that the posterior usually still contains a mode close to the true composition, even when it is not the highest-weight one.

A direct accuracy comparison remains out of scope (\autoref{sec:related}), but published refits give a sense of scale. \citet{merlinSurfaceCompositionPhysical2010} fit six TNOs with both Hapke and Shkuratov and find single-component proportions differing by up to ${\sim}0.28$ between models, with grain sizes spanning an order of magnitude; \citet{barucci50000QuaoarSurface2015} fit four Shkuratov spectra of Quaoar and find proportions varying by up to ${\sim}0.13$ and grain sizes by ${\sim}0.2$--$0.6$\,dex across epochs.

\section{Limitations and Future Work}

\textbf{Novel components.} The OOD split tests unseen combinations of the same $K=20$ components. TNFlow cannot recover a material it has never seen. The model can be retrained with any component set, however, and retraining is a one-time offline cost.

\textbf{Generalization to real objects.} A qualitative analysis (\autoref{app:jwst}) applies TNFlow to real spectra from the JWST DiSCo-TNOs survey \citep{pinilla-alonsoJWSTDiSCoTNOsPortrait2025}. CO$_2$, CO, and CH$_3$OH, whose absorption bands are clearly visible in the spectra, are not fully recovered by the model. Meanwhile, H$_2$O is assigned as the dominant component for all objects, including those the survey classifies as water-poor. This behavior was not seen on synthetic spectra. We hypothesize this gap is due to model bias against trace-abundance components (\autoref{app:trace}), or because the Shkuratov RTM does not reproduce all physical effects present in real observations.

\textbf{Future work.} To address the sim-to-real gap, we plan to curate a training set that better covers trace abundances and other underrepresented cases in the current corpus. Creating hybrids of real and synthetic spectra may also help the model learn physical effects that the simulator does not capture.

To overcome the fixed component set, we will explore returning compositions over arbitrary-length optical-constant profiles rather than a set list of $K$ named materials. Finally, resynthesizing each recovered mode through the RTM and comparing it to the input spectrum would verify if all modes are degenerate solutions.

In summary, TNFlow demonstrates that amortized posterior inference over TNO surface compositions is practical: a sub-second, fully offline surrogate for the Shkuratov RTM that replaces the manual hours-per-object fitting loop and makes the inherent degeneracy of the inverse problem explicit. On synthetic data, composition accuracy is comparable to or better than the disagreements typically reported between independent manual refits. The main open challenge is closing the sim-to-real gap on real JWST spectra, which will require improvements to both training-set coverage and simulator fidelity.

\begin{ack}
This research was funded and made possible by the SETI Institute Research Experience for Undergraduates (REU) program, led by Dr. Matthew S.\ Tiscareno. We thank Dr. Cristina M.\ Dalle Ore for her REU mentorship and for providing the bestfit spectral collection that forms the foundation of the training corpus used in this work. We thank the SETI Institute for the computational resources used to train the model. We thank Dr. Alessandra Ricca's group at NASA Ames, supported by NASA Solar System Workings grant 80NSSC21K0776. We thank Dr. Jeffery Smith for useful conversations and comments on the paper draft.
\end{ack}

\bibliographystyle{plainnat}
\bibliography{tno-ml}

\newpage
\appendix
\section{Training Corpus}
\label{app:components}

The training corpus of 463,275 synthetic Shkuratov spectra is assembled from two sources.

\textbf{Pre-existing bestfit collection (163,285 spectra).} A library of pre-computed Shkuratov intimate-mixture and pure-component spectra spanning 17 of the 20 components, with mixtures of 1--4 components and varied grain sizes. These spectra were generated in prior work using the Shkuratov model with published laboratory optical constants; the full generation parameters and optical-constant sources for each spectrum are available in the companion repository.

\textbf{Purpose-built extension (299,990 spectra).} To increase the diversity and size of the corpus, we generated additional spectra using CANA \citep{cana2018} and SDOC\footnote{\url{https://github.com/cana-asteroids/sdoc}}, two open-source tools for computing Shkuratov reflectance from optical constants. This extension covers 14 components in mixtures of 2--7 components (50,000 spectra per mixture size from 2 to 7), with compositions sampled uniformly on the simplex and grain sizes sampled log-uniformly. It adds three components not present in the pre-existing data (CO, C$_2$H$_4$, pyroxene) and substantially increases the representation of trace-abundance ($<$5\%) component slots.

\autoref{tab:components} lists all $K{=}20$ components and the number of spectra containing each, split by source.

\begin{table}[h]
  \caption{Component library ($K{=}20$) and per-component spectrum counts. ``Bestfit'' is the pre-existing collection; ``Generated'' is the CANA/SDOC extension. Optical-constant references are given for the generated extension; bestfit-collection sources are documented in the companion repository.}
  \label{tab:components}
  \centering
  \scriptsize
  \begin{tabular}{llrrrL{3.6cm}}
    \toprule
    Component & Formula & Bestfit & Gen. & Total & OC Reference \\
    \midrule
    Amorphous carbon (AC)    & C                        & 72,991  & 107,795 & 180,786 & \citet{rouleauShapeClusteringEffects1991} \\
    Titan tholin             & ---                      & 57,841  & 106,883 & 164,724 & \citet{khareOpticalConstantsOrganic1984} \\
    Olivine                  & (Mg,Fe)$_2$SiO$_4$      & 43,140  & 107,526 & 150,666 & \citet{dorschnerStepsInterstellarSilicate1995} \\
    Triton tholin            & ---                      & 43,130  &  95,504 & 138,634 & \citet{khareOpticalConstantsTriton1994} \\
    H$_2$O                   & H$_2$O                   & 32,928  & 103,204 & 136,132 & \citet{mastrapaOpticalConstantsAmorphous2008,mastrapaOPTICALCONSTANTSAMORPHOUS2009} \\
    CH$_3$OH (methanol)      & CH$_3$OH                 & 30,715  &  96,553 & 127,268 & \citet{hudginsMidFarInfraredSpectroscopy1993} \\
    CO$_2$                   & CO$_2$                   & 19,690  &  94,336 & 114,026 & \citet{hansenSpectralAbsorptionSolid1997} \\
    CH$_4$ (methane)         & CH$_4$                   & 18,633  &  94,493 & 113,126 & \citet{grundyTemperatureDependentSpectrumMethane2002} \\
    Pyroxene                 & (Mg,Fe)SiO$_3$           &      0  & 105,049 & 105,049 & \citet{dorschnerStepsInterstellarSilicate1995} \\
    N$_2$                    & N$_2$                    &  6,185  &  93,868 & 100,053 & \citet{schmittOpticalPropertiesIces1998} \\
    CO                       & CO                       &      0  &  94,137 &  94,137 & \citet{pubchemCarbonMonoxide} \\
    C$_2$H$_4$ (ethylene)    & C$_2$H$_4$               &      0  &  86,226 &  86,226 & \citet{hudsonInfraredSpectraOptical2014} \\
    C$_2$H$_6$ (ethane)      & C$_2$H$_6$               &    252  &  82,760 &  83,012 & \citet{hudsonInfraredSpectraOptical2014} \\
    NH$_3$ (ammonia)         & NH$_3$                   &     30  &  81,645 &  81,675 & \citet{roserInfraredComplexRefractive2021} \\
    PM100                    & ---                      & 33,765  &       0 &  33,765 & --- \\
    NaCl                     & NaCl                     & 19,117  &       0 &  19,117 & --- \\
    CO in N$_2$              & CO:N$_2$                 & 18,455  &       0 &  18,455 & --- \\
    Ice Tholin II            & ---                      & 14,505  &       0 &  14,505 & --- \\
    Serpentine               & ---                      & 11,265  &       0 &  11,265 & --- \\
    HAC                      & C:H                      &  9,075  &       0 &   9,075 & --- \\
    \bottomrule
  \end{tabular}
\end{table}

\section{Application to Real JWST Spectra}
\label{app:jwst}

\begin{figure}[h]
  \centering
  \includegraphics[width=\linewidth]{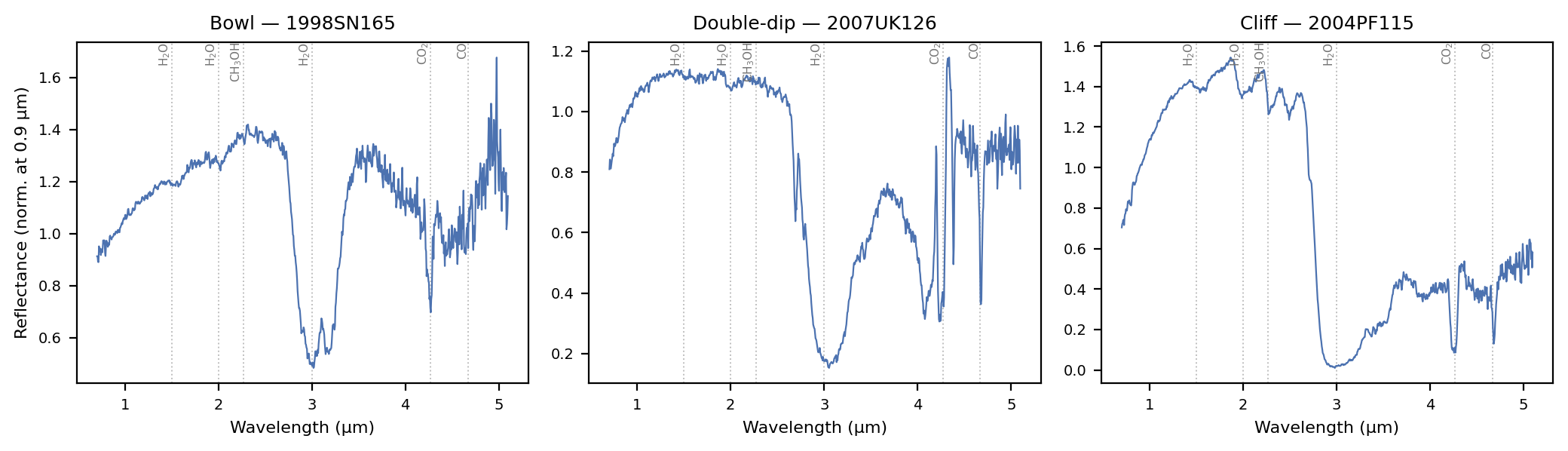}
  \caption{The three DiSCo-TNOs exemplars, normalized at 0.9\,\textmu m. Dotted lines mark
           the bands discussed in the text.}
  \label{fig:jwst}
\end{figure}

We show results of TNFlow inference on a random representative of each DiSCo-TNOs spectral group \citep{pinilla-alonsoJWSTDiSCoTNOsPortrait2025} (\autoref{fig:jwst}). TNFlow inference used 500 draws and the settings of \autoref{sec:results}. The spectra cover a smaller range of wavelength than the training data, but every band of interest is present.

\begin{table}[h]
  \caption{All modes recovered for the three exemplars. Components below 0.01 proportion are omitted.}
  \label{tab:jwst}
  \centering
  \footnotesize
  \begin{tabular}{llll}
    \toprule
    Object (group) & Mode & $w$ & Composition \\
    \midrule
    1998\,SN165          & 1 & 0.49 & AC 0.71, H$_2$O 0.22, NaCl 0.07 \\
    (bowl)               & 2 & 0.39 & AC 0.57, H$_2$O 0.23, NaCl 0.09, TTh 0.08 \\
                         & 3 & 0.12 & AC 0.39, H$_2$O 0.22, Oliv 0.16, NaCl 0.12, TTh 0.07, NH$_3$ 0.02 \\
    \midrule
    2007\,UK126          & 1 & 0.71 & H$_2$O 0.39, Oliv 0.17, AC 0.14, NH$_3$ 0.11, CO$_2$ 0.10, TTh 0.09 \\
    (double-dip)         & 2 & 0.20 & H$_2$O 0.40, Oliv 0.25, NH$_3$ 0.14, AC 0.11, CO$_2$ 0.09 \\
                         & 3 & 0.09 & H$_2$O 0.40, AC 0.21, NH$_3$ 0.17, CO$_2$ 0.12, TTh 0.04, NaCl 0.03 \\
    \midrule
    2004\,PF115          & 1 & 0.65 & H$_2$O 0.69, TTh 0.30 \\
    (cliff)              & 2 & 0.33 & H$_2$O 0.65, TTh 0.30, CH$_3$OH 0.05 \\
                         & 3 & 0.03 & H$_2$O 0.67, TTh 0.22, IT2 0.05, AC 0.04, CH$_3$OH 0.01 \\
    \bottomrule
  \end{tabular}
\end{table}

CO$_2$ is found in all solutions for 2007\,UK126, where its 4.27\,\textmu m band is deep and pronounced. However, it is missed on the bowl and cliff where the band is still present. CO bands are also present in all three spectra, but it never appears in a solution. CH$_3$OH is clearly present in 2004\,PF115, but is not found in the top mode. H$_2$O is assigned dominantly to all three objects, including on the double-dip and cliff, which \citet{pinilla-alonsoJWSTDiSCoTNOsPortrait2025} describe as water-poor.

However, the model gets many things consistent with the survey as well. The bowl-type object is amorphous-carbon dominated, Titan Tholin is found in the cliff-type object, and Olivine is plausible on the double-dip-type object.

\section{Calibration by Abundance}
\label{app:trace}

\begin{figure}[h]
  \centering
  \includegraphics[width=\linewidth]{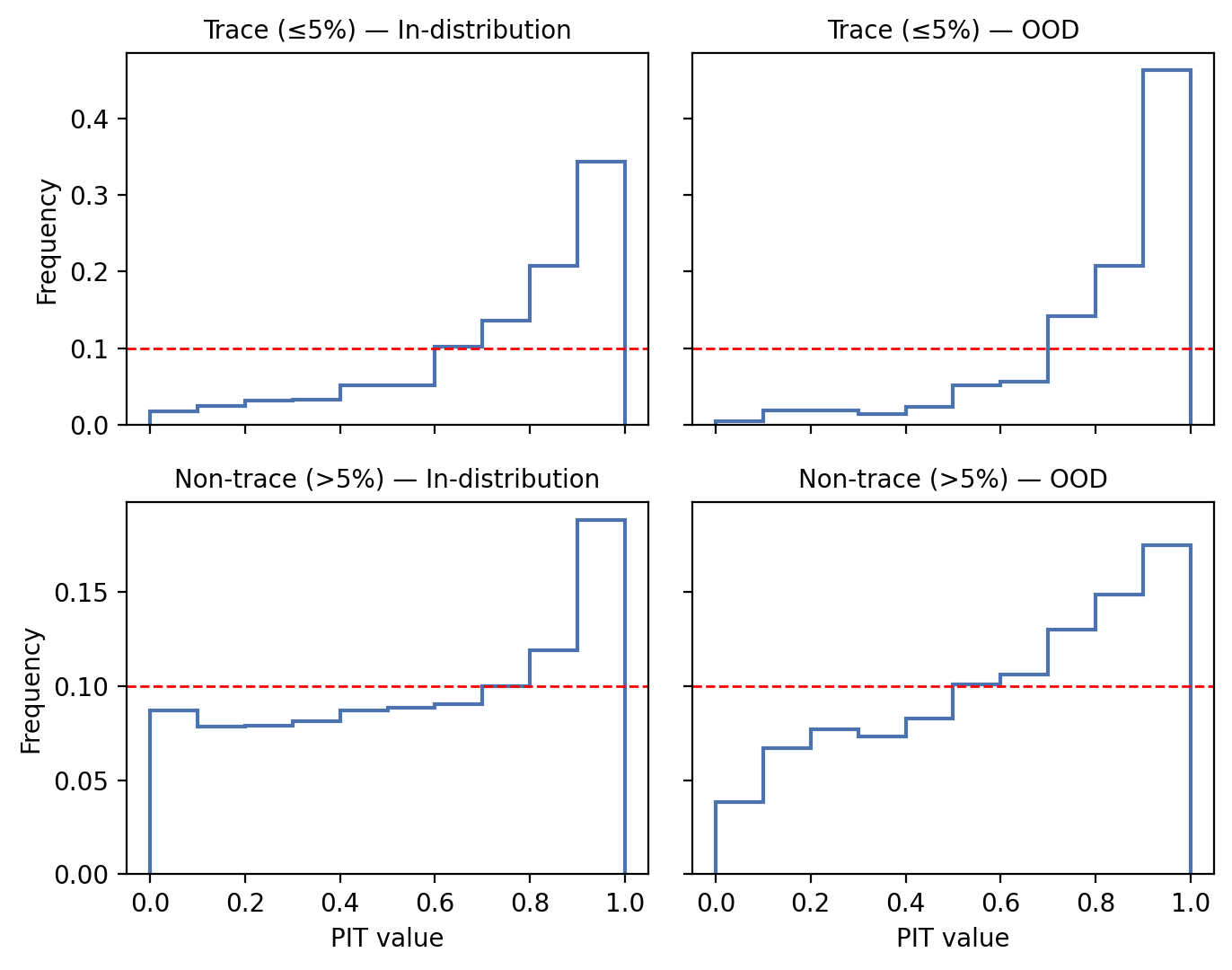}
  \caption{Composition PIT split by target proportion: trace ($\leq$0.05, top) and
           non-trace ($>$0.05, bottom), on the test (left) and OOD (right) splits.
           A uniform histogram (dashed) indicates calibration.}
  \label{fig:pit_trace}
\end{figure}

Splitting the composition PIT of \autoref{fig:pit} at 0.05 proportion shows that the miscalibration disproportionately affects components with trace abundance (\autoref{fig:pit_trace}). On the test split the final PIT bin holds $1.9{\times}$ the uniform expectation for components above 0.05 proportion, but $3.4{\times}$ for those at or below it; on the OOD split the same figures are $1.8{\times}$ and $4.7{\times}$. The systematic underprediction reported in \autoref{sec:discussion} is more concentrated in trace components. We believe this may be due to the training corpus not densely covering spectra with trace components.

\section{Noise Augmentation}
\label{app:noise}

Synthetic spectra are augmented with a fresh draw each time they are batched. For a spectrum with
reflectance $R(\lambda)$ we draw two per-spectrum scales $\sigma_0, a \sim \mathcal{N}(0, 0.04)$ and
per-channel standard normals $u(\lambda), v(\lambda) \sim \mathcal{N}(0,1)$, and set
\begin{equation*}
  R'(\lambda) \;=\; R(\lambda)\left[\,1 + \sigma_0\, u(\lambda) \sqrt{\lambda / 0.9\,\mu\mathrm{m}}\,\right]
              \;+\; a\, v(\lambda)\, \mathrm{median}(R).
\end{equation*}
The multiplicative term grows as $\sqrt{\lambda}$; the additive term is a flat floor set by the
spectrum's own median reflectance. Since $u$ and $v$ are symmetric, only the magnitudes of $\sigma_0$
and $a$ matter, so the effective per-spectrum scales are half-normal with scale $0.04$. Augmentation
is applied to synthetic rows only: the real spectra of \autoref{app:jwst} are passed through
unmodified.

These levels were chosen ad hoc, as a rough stand-in for how noise appears in real spectra rather
than a calibration against a specific instrument noise model.

\section{Comparison Scope}
\label{app:comparison}

MCMC inversion of the Hapke RTM \citep{schmidtRealisticUncertaintiesHapke2015} is automated and returns a posterior, so an accuracy comparison is meaningful in principle. However, two obstacles place it outside this work: that method targets the Hapke RTM rather than Shkuratov, and it returns Hapke parameters and optical constants where TNFlow returns abundances over a fixed component set. Equating the two output spaces requires methodology beyond our scope. Manual fitting and linear unmixing return point estimates, which cannot be set against a multimodal posterior on the terms that matter here.

\section{Compute Resources}
\label{app:compute}

\textbf{Training.} All training was performed on a single NVIDIA RTX 6000 Ada Generation GPU (48\,GB VRAM) on a shared workstation (64-core Xeon, 768\,GB RAM, Ubuntu).

\textbf{Inference.} The per-object timing reported in Section~\ref{sec:results} (${\sim}0.7$\,s, 500 draws + BIC-GMM) was measured on a laptop with an Intel Core i5-8257U at 1.40\,GHz and 8\,GB RAM, using CPU only.

\end{document}